\documentclass[%
 reprint,
 longbibliography,
 amsmath,
 amssymb,
 aps,
 pre,
]{revtex4-1}

\usepackage{dcolumn}
\usepackage{bm}

\usepackage{amssymb}
\usepackage{amsmath}
\usepackage{cases}
\usepackage{graphicx,color}
\usepackage{array}
\usepackage{multirow}
\usepackage{makecell}
\usepackage{mathtools}

\definecolor{r}{rgb}{1,0,0}   
\definecolor{g}{rgb}{0,1,0}   
\definecolor{b}{rgb}{0,0,1}
\definecolor{purple}{rgb}{0.808,0.454,0.718}

\begin{document}


\title{Average Evolution and Size-Topology Relations for Coarsening 2d Dry Foams}


\author{Anthony T. Chieco$^1$, James P. Sethna$^2$ and Douglas J. Durian$^{1}$}
\affiliation{$^1$Department of Physics and Astronomy, University of Pennsylvania, Philadelphia PA, USA\\
$^2$Department of Physics, Cornell University, Ithaca NY, USA}


\date{\today}

\begin{abstract}
Two-dimensional dry foams coarsen according to the von~Neumann law as $dA/dt \propto (n-6)$ where $n$ is the number of sides of a bubble with area $A$. Such foams reach a self-similar scaling state where area and side-number distributions are stationary. Combining self-similarity with the von~Neumann law, we derive time derivatives of moments of the bubble area distribution and a relation connecting area moments with averages of the side-number distribution that are weighted by powers of bubble area. To test these predictions, we collect and analyze high precision image data for a large number of bubbles squashed between parallel acrylic plates and allowed to coarsen into the self-similar scaling state. We find good agreement for moments ranging from two to twenty.
\end{abstract}



\maketitle




\section{Introduction}

Dry two-dimensional foams can be made by squashing bubbles between parallel plates and letting most of the liquid drain out. Viewed from normal to the plates, neighboring bubbles are separated by thin soap films that are circular arcs and that -- according to Plateau's Laws -- meet at $120^\circ$ at 3-fold vertices~\cite{WeaireRivier1984, WeaireHutzlerBook, Cantat2013FoamsBook, Langevin2020}. As a consequence, Euler's rule implies that the average number of sides is $\langle n\rangle=6$ if the sample is sufficiently large. It is commonly observed that bubbles with more sides $n$ tend to be bigger. Various size-topology relations have been proposed to quantify this behavior~\cite{Rivier85, ChiuReview}.  One of the oldest is Lewis's Law~\cite{Lewis1930}, which states that the area of $n$ sided-bubbles is a linear function of $n$.  Desch's Law analogously holds that the perimeter of $n$-sided bubbles is linear in $n$. These ``laws" are actually empirical approximations, found to hold to varying degrees and with different linear relations for different kinds of cellular structures, \emph{e.g.}\ plant and animal tissues or foams or grains in an alloy. For foams, systematic deviations from the Lewis and Desch laws were recently observed and accounted for by a simplified ``granocentric" model \cite{BrujicPRL12} in which a central particle is uniformly surrounded by $n$ equidistant bubbles of the same size \cite{RothPRE2013}.

Even in the absence of soap film rupture and drainage of the liquid between bubbles, foams coarsen with time. Locally, gas diffuses between neighboring bubbles according to their pressure difference.  This tends to make small bubbles shrink and large bubbles grow.  Based on little more than geometry, von~Neumann's law~\cite{vonNeumannLaw, Mullins56, Stavans93} states that the area $A_i$ of a bubble $i$ with $n_i$ sides changes at rate
\begin{equation}
    \frac{d A_i}{dt} = K_o(n_i-6)
\label{eq_vonNeumann}
\end{equation}
where $K_o$ is a materials constant proportional to surface tension, the solubility and diffusivity of the gas, and inversely proportional to film thickness. This law is exact, no matter what the sizes and shapes of the neighboring bubbles. If an initial foam sample is not too pathological or small, it will evolve into a statistically self-similar scaling state where the side-number distribution and dimensionless ratios of bubble area moments are all independent of time~\cite{Mullins1986} (see Refs.~\cite{GlazierGrossStavans87, MarderPRA87, BeenakkerPRA88, WeaireLei1990, GlazierAndersonGrest90, HerdtleArefJFM} on the approach to the scaling state, and Refs.~\cite{Stavans90, RothPRE2013} on the demonstration of scaling, for dry 2d foams). In such a state, the von~Neumann law can be used to show that the average coarsening rate is constant and equal to
\begin{equation}
    \frac{d\langle A\rangle}{dt} = 2K_o\frac{\langle A\rangle^2}{\langle A^2\rangle} \big[\langle\langle n\rangle\rangle-6 \big]
\label{eq_avgvonNeumann}
\end{equation}
where $\langle A\rangle = (\sum_{i=1}^N A_i)/N = A_{tot}/N$ is the average bubble size, $N$ is the total number of bubbles, $A_{tot}$ is the sample area, $\langle A^2\rangle = (\sum_{i=1}^N {A_i}^2)/N$, $\langle\langle n\rangle\rangle = \sum n F(n)$ is the area-weighted average side number, and $F(n)$ is the area-weighted side number distribution -- \emph{i.e.} the fraction of sample area inside $n$-sided bubbles~\cite{RothPRE2013}. While the probability $P(n)$ that a randomly-chosen \emph{bubble} is $n$-sided has been widely studied, the probability $F(n)$ that a randomly chosen \emph{point} is inside an $n$-sided bubble is more directly important for the average coarsening behavior. According to Eq.~(\ref{eq_avgvonNeumann}), the average coarsening rate depends on both bubble sizes, through the moment ratio $\langle A^2\rangle / \langle A\rangle^2$, as well as on topology, through $\langle\langle n\rangle\rangle$.

In this paper, we generalize three ways upon Eq.~(\ref{eq_avgvonNeumann}), and we use one of the results to predict a generalized size-topology relation.  Then we make experimental tests.

\section{Predictions}

To assist with derivations, we use single angle brackets to denote numeric averages and double angle brackets to denote weighted averages. Specifically, we write the average $p$-th power of bubble area as
\begin{equation}
\langle A^p \rangle = \frac{1}{N}\sum_{i=1}^N {A_i}^p
\label{eq_avgAp}
\end{equation}
And we write the $A^p$-weighted average side numbers as
\begin{equation}
    \langle\langle n\rangle\rangle_p = \frac{ \sum_{i=1}^N n_i {A_i}^p }{\sum_{i=1}^N {A_i}^p }
\label{eq_npavg}
\end{equation}
Note that $\langle\langle n\rangle\rangle_1$ is the area-weighted average side number $\langle\langle n\rangle\rangle$ used in Eq.~(\ref{eq_avgvonNeumann}), and $\langle\langle n\rangle\rangle_0 = \langle n\rangle = 6$ is the familiar numeric average size number.

For our first generalization of Eq.~(\ref{eq_avgvonNeumann}), we begin by writing and rearranging the following identity:
\begin{eqnarray}
\langle A\rangle^p &=& \langle A\rangle^p \times \frac{\langle A\rangle}{\langle A\rangle} \times \frac{\sum {A_i}^p}{\sum {A_i}^p} \times \frac{\sum {A_i}^{p+1}}{N\langle A^{p+1}\rangle}
\label{eq_identity1} \\
 &=& \frac{1}{N\langle A\rangle} \times  \frac{ \langle A\rangle^{p+1}}{\langle A^{p+1}\rangle}  \times \frac{\sum {A_i}^p}{\sum {A_i}^p} \times \sum {A_i}^{p+1}
\label{eq_identity2}
\end{eqnarray}
where the sums run over all bubbles, from $i=1$ to $i=N$.  Note that the first three terms in Eq.~(\ref{eq_identity2}) are all independent of time: The first is the reciprocal of sample area, the second is constant when the sample is in a self-similar scaling state, and the third equals one. Therefore, it is straightforward to differentiate both sides with respect to time. Using von~Neumann's law on the fourth term on the right hand side, then recognizing $\sum {A_i}^p/N=\langle A^p\rangle$ and rearranging, gives
\begin{equation}
    p\langle A\rangle^{p-1}\frac{d\langle A\rangle}{dt} = \frac{\langle A\rangle^p\langle A^p\rangle}{\langle A^{p+1}\rangle} \frac{\sum (p+1){A_i}^p K_o(n_i-6)}{\sum {A_i}^p}
\end{equation}
which simplifies to the following generalization of Eq.~(\ref{eq_avgvonNeumann}):
\begin{equation}
    \frac{d\langle A\rangle}{dt} = K_o\left( 1+\frac{1}{p}\right) \frac{\langle A\rangle\langle A^p\rangle}{\langle A^{p+1}\rangle} \big[ \langle\langle n\rangle\rangle_p - 6 \big]
\label{eq_dAdt}
\end{equation}
This holds for any nonzero value of $p$, not just integers. Near $p=0$ it implies $d\langle A\rangle/dt = K_o \lim_{p \to 0}[\langle\langle n\rangle\rangle_p-6]/p$, which proves $\langle n\rangle=6$ for steady state without use of Euler's rule.  Since the left hand side of Eq.~(\ref{eq_dAdt}) is the same for any value of $p$, we may equate the right hand sides evaluated at $p$ and at $p\rightarrow q$. This gives the final result
\begin{equation}
    \frac{p(q+1)\langle A^{p+1}\rangle\langle A^q\rangle}{q(p+1)\langle A^p\rangle\langle A^{q+1}\rangle} = \frac{\langle\langle n\rangle\rangle_p - 6}{\langle\langle n\rangle\rangle_q - 6}
\label{eq_pq}
\end{equation}
relating bubbles sizes, on the left, and network topology, on the right, which holds in the self-similar scaling state.  It will be tested experimentally in later sections for the special case $q=p-1$, where it can be rewritten as
\begin{equation}
    \frac{ \langle A^p\rangle^2 }{\langle A^{p-1}\rangle \langle A^{p+1}\rangle} \frac{ \langle\langle n\rangle\rangle_p-6 }{ \langle\langle n\rangle\rangle_{p-1}-6} = \frac{p^2}{p^2-1}
\label{eq_sizetopo}
\end{equation}
Here, the product of size and topology ratios decrease towards one as $p$ increases towards infinity. As per Eq.~(\ref{eq_dAdt}), it holds for general values of $p$. Note that both sides vanish in the limit $p\rightarrow 0$ and both sides diverge in the limit $p\rightarrow 1$ owing to $\langle\langle n\rangle\rangle_0 \equiv \langle n\rangle = 6$.  We note, too, that Eq.~(\ref{eq_pq}) can be derived from Eq.~(\ref{eq_sizetopo}).

For a second generalization of Eq.~(\ref{eq_avgvonNeumann}), we consider the time derivative of $\langle A^p\rangle = \frac{1}{N}\sum {A_i}^p$. Substituting $1/N=\langle A\rangle/A_{tot}$ and using the product differentiation rule, then von~Neumann's law, gives
\begin{eqnarray}
    \frac{d\langle A^p\rangle}{dt} &=& \frac{1}{A_{tot}}\frac{d\langle A\rangle}{dt}\sum {A_i}^p + \frac{\langle A\rangle}{A_{tot}}\frac{d}{dt}\sum {A_i}^p
    \label{eq_dApdt1} \\
    &=& \frac{\langle A^p\rangle}{\langle A\rangle}\frac{d\langle A\rangle}{dt} + \frac{1}{N}\sum p{A_i}^{p-1}K_o(n_i-6)
    \label{eq_dApdt2} \\
    &=& \frac{\langle A^p\rangle}{\langle A\rangle}\frac{d\langle A\rangle}{dt} + pK_o\langle A^{p-1}\rangle\big[ \langle\langle n\rangle\rangle_{p-1} - 6\big]
    \label{eq_dApdt3} 
\end{eqnarray}
For $p=1$ these expressions reduce to $d\langle A\rangle/dt = d\langle A\rangle/dt$ and hence give nothing new.  For $p>1$ we may simplify further by assuming that the foam is in a self-similar scaling state, which permits $d\langle A\rangle/dt$ to be evaluated with Eq.~(\ref{eq_dAdt}) at $p\rightarrow p-1$. This gives the final result:
\begin{equation}
    \frac{d\langle A^p\rangle}{dt} = K_o\frac{p^2}{p-1} \langle A^{p-1}\rangle \big[ \langle\langle n\rangle\rangle_{p-1} -6 \big]
\label{eq_dAp}
\end{equation}
which grows in proportion to $t^{p-1}$ and holds for $p>1$.

For the third and perhaps prettiest generalization of Eq.~(\ref{eq_avgvonNeumann}), we compute the time derivative of $\langle A^{p+1}\rangle / \langle A^p\rangle = \sum {A_i}^{p+1} / \sum {A_i}^p$.  This quantity has units of area, so in the scaling state its time derivative should be constant.  Indeed, using the quotient rule with von~Neumann's law, then simplifying with Eq.~(\ref{eq_sizetopo}), gives
\begin{equation}
    \frac{d}{dt}\frac{ \langle A^{p+1}\rangle }{ \langle A^p\rangle} = K_o\left( 1+\frac{1}{p} \right) \big[ \langle\langle n\rangle\rangle_p - 6 \big]
\label{eq_DAppOverAp}
\end{equation}
This result can be alternatively derived using Eq.~(\ref{eq_dAp}). To test if the foam is in a scaling state, experimental data for average bubble growth should collapse to a line of slope $K_o$ if plotted as $\langle A^{p+1}\rangle / \langle A^p\rangle$ versus $(1+1/p) [\langle\langle n\rangle\rangle_p - 6] t$ for several values of $p$.

It might be interesting to similarly compute the time derivative of quantities such as $\langle A^{p+q+1}\rangle / [\langle A^p\rangle \langle A^q \rangle]$ or $\langle A^{pq+1} \rangle / \langle A^p\rangle^q$, which have units of area and hence are also constant in the scaling state. Note that Eq.~\ref{eq_sizetopo} gives us one equation for every two unknowns. If we could get another size-topology relation for every $p$, it might be possible to combine with the above identities to solve for $\langle A^p\rangle / \langle A\rangle^p$ and $\langle\langle n\rangle\rangle_p$, and
conceivably derive (for example) an analytic form for the bubble size distribution.

\section{Materials and Methods}

To test the predictions by experiment we generate aqueous foams of Nitrogen bubbles that have no film rupture and that, as we will show, coarsen in a self similar scaling state. These foams are made from a solution that is 92\% deionized water and 8\% Dawn Ultra Concentrated dish detergent. The foam is generated inside a sample cell constructed from two 1.91~cm-thick acrylic plates separated by a spacing $H=0.21$~cm and sealed with two concentric o-rings, the inner of which has a 23~cm diameter; this is the same apparatus used in \cite{RothPRE2013, ATCcoarsen}, where additional details may be found. 
 
Foams are produced as follows. First the cell is filled completely with foaming solution. It is then flushed with Nitrogen and sealed when a desired amount of liquid remains. The entire sample cell is vigorously shaken for several minutes until the gas is uniformly dispersed as bubbles that are smaller than the gap between the plates. This ensures a large number of small bubbles and aids in repeatability, tested later. The foam is thus initially very wet, opaque, and three-dimensional. We stand the cell so that the plane of the foam is vertical and place it between a Vista Point~A light box and a Nikon~D850 camera with a Nikkor AF-S 300mm 1:2.8D lens. Images are acquired every 5~minutes for up to 24~hours. Several minutes after production most of the liquid drains out of the foam; after an hour the bubbles become large compared to the gap. The resulting foam is dry and quasi- two-dimensional, and only subsequent data are kept. In this regime, the radius of the Plateau borders is about 0.03~mm, so the thin soap films span more than 97\% of the gap $H$ between the plates and gas transport across the borders is negligible. 
 
The areas of individual bubbles are found directly from the images. The images undergo some some slight post processing to enhance contrast and then we binarize, skeletonize, and watershed them. Bubbles are the watershedding basins of the skeletonized images and the number of pixels within each basin is converted into the bubble area. Thus the choice of camera is important and the D850 has an $8256 \times 5504~{p_o}^2$ pixel array where $p_o$ is the pixel size; combining it with a telephoto lens placed $1.5$~m away allows us to have both a large number of bubbles and an accurate measurement of their area.

The vertices of each bubble are also found from the watershed image as the pixels where three basins are in contact. Each of the three bubbles are assigned that vertex and the total number of vertices of each bubble equals its number of sides. We track the number of sides of each bubble and find they only change when there is a topological rearrangement; thus we verify there are no film ruptures throughout the experiment. Only bubbles that do not overlap the boundary of the $19\times 6.5$~cm$^2$ region of interest are kept for analysis.  With the resulting collection of relevant size and topology information for bubbles from each image we can now test predictions from the previous section.

\section{Experimental Results and Discussion}

We are interested in examining the predictions made by two generalizations of Eq.~\ref{eq_avgvonNeumann}. The first is Eq.~\ref{eq_DAppOverAp} which uses distributions of the bubble areas and sides to predict a coarsening rate for the entire foam. And the second is Eq.~\ref{eq_sizetopo} which is a generalized size-topology relationship. These equations were derived for 2-dimensional foams in a scaling state so we first establish the foam is self-similar. Once this is determined, we test the expectations against the results garnered from three separate experiments. To distinguish the samples and gauge repeatability we identify them as Foam A, B, and C. Each foam is produced using the method stated in the previous section; the experiments differ by the specific initial distributions of 3-dimensional bubble sizes and the dates they were performed; the data for Foam C was collected several months before the data for Foam A and B.  We keep images only when the foams have become quasi-2d. Fig.~\ref{foam_rescale} show representative zoomed-in images, where the bubbles appear as polygons with curved edges. The three data sets consist of $\{81,101,69\}$ images for the three samples, respectively, as they coarsen: the corresponding initial numbers of bubbles are $\{1745,3025,2588\}$, and the final numbers of bubble are $\{728,688,846\}$; the corresponding initial and final average areas of the bubbles are $\{6.7,3.5,4.4\}$~mm$^2$, and $\{15.6,14.4,12.7\}$~mm$^2$, respectively;  In supplementary material we provide all bubble area and side number data sorted by time for the three experiments.

\subsection{Self-Similar Scaling State}

\begin{figure}[h]
\includegraphics[width=3.5in]{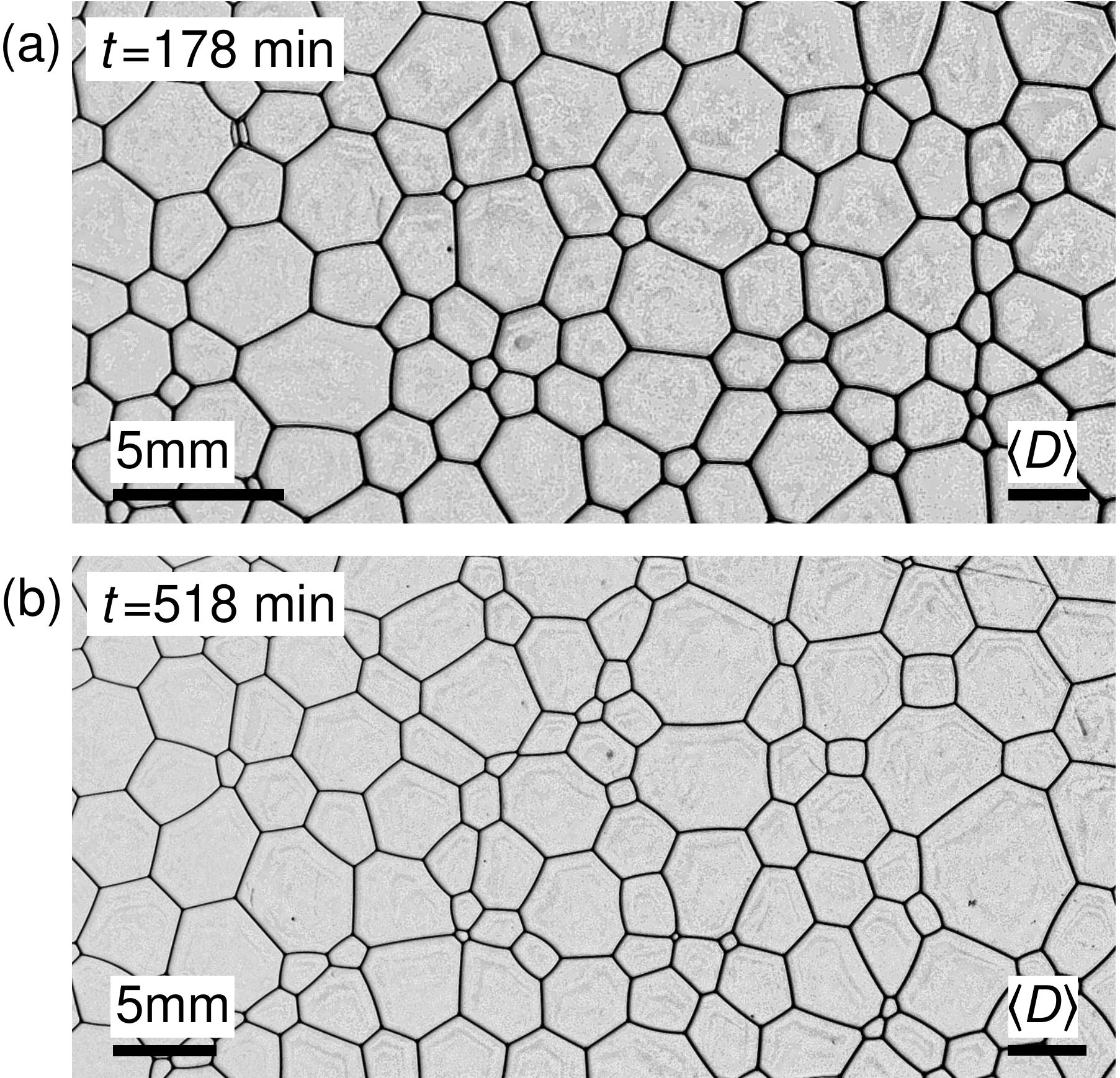}
\caption{Images of coarsening foam at different times, zoomed in so that the distributions of sizes appear the same by eye. Here, the image side lengths are both $L_x=15\sqrt{\langle A\rangle}$ and $L_y=7.5\sqrt{\langle A\rangle}$. This is a qualitative demonstration of self-similarity in the scaling state.}
\label{foam_rescale}
\end{figure}

Self-similarity is well documented and has been observed in experiment and simulation.  Here, a qualitative demonstration is illustrated by Fig.~\ref{foam_rescale}. There, we show two images separated by 340~minutes and zoom in to a portion from each that is $\left[L_x,L_y\right]=[15,7.5] \sqrt{\langle A \rangle}$, so that the average bubble occupies the same visual space in each.  Bubbles in the younger foam are smaller and thus appear more pixelated.  Otherwise the two images appear very similar, as though coming from different regions of the same sample and hence as having the same apparent bubble size distributions.

\begin{figure}[h]
\includegraphics[width=3.5in]{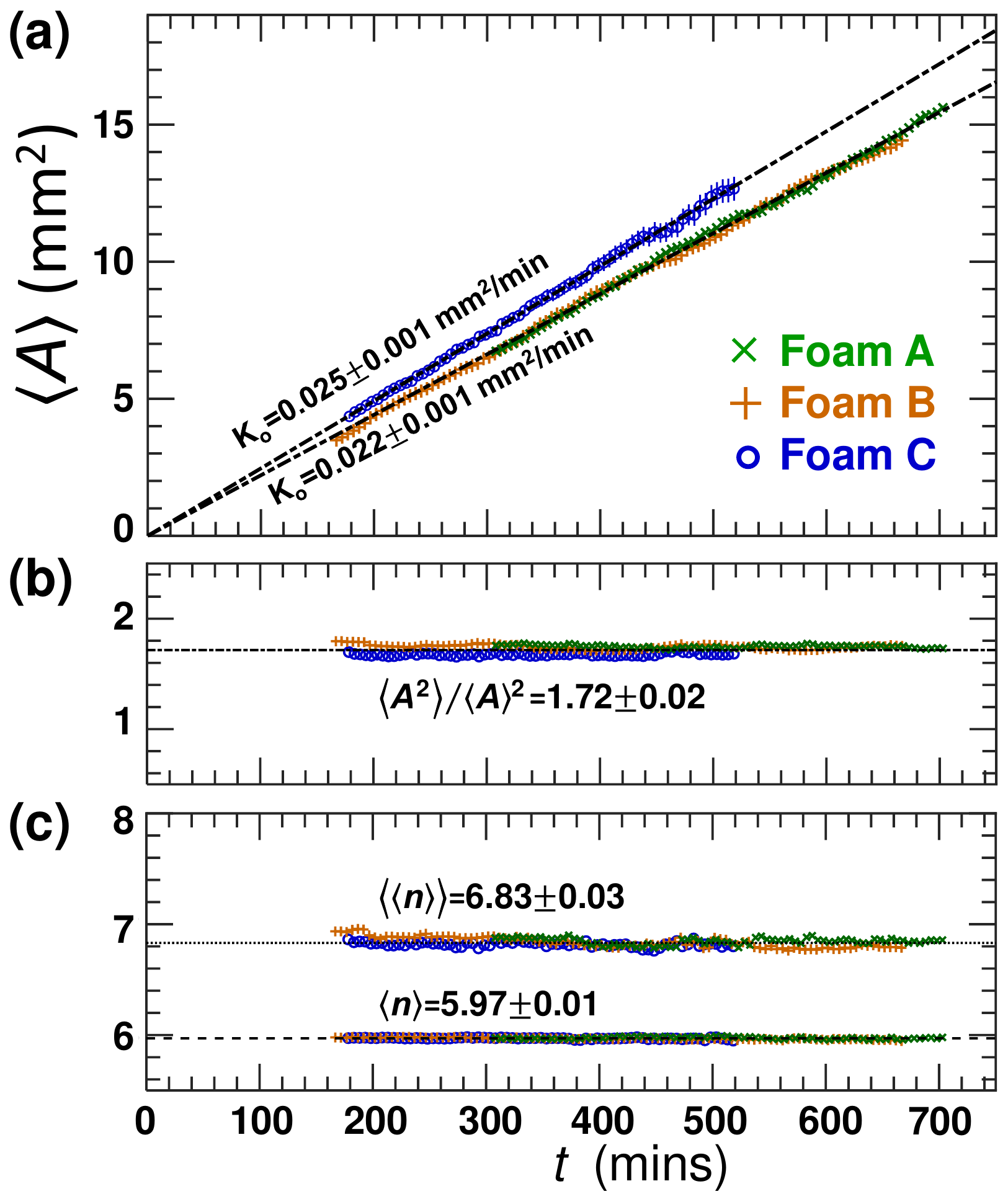}
\caption{(a) Average area, (b) second moment divided by the average area squared, and (c) area-weighted average side number $\langle\langle n\rangle\rangle$ as well as the average side number $\langle n\rangle$ versus time. The different symbols represent data from the three different foam samples. In part (a) statistical error bars are included for Foam C as $\sigma_A=\langle A\rangle/\sqrt{N}$ where $N$ is the number of bubbles.  The statistical uncertainties for the other samples are comparable, and correspond well to the bumps and wiggles in the data and also to the differences between samples A and B. Also in (a) the dot-dashed represent proportionality fits, giving $K_o$ values as labeled.}
\label{scaling_state_data}
\end{figure}

For a quantitative demonstration that the foams are in a scaling state, we consider several metrics. Fig.~\ref{scaling_state_data}(a) shows that the average area increases linearly with time for all three of our foams. Such linearity holds only in the self-similar regime (too-monodisperse and too-polydisperse sample have initial coarsening rates that are respectively slower or faster~\cite{GlazierGrossStavans87, BeenakkerPRA88}). Furthermore, this linear behavior is predicted by Eq.~\ref{eq_avgvonNeumann} for a self similar foam because both the ratio $\langle A^2 \rangle / \langle A \rangle^2$ and the area-weighted average side number $\langle \langle n \rangle \rangle$ are independent of time. These quantities are plotted in parts (b) and (c), respectively; both are constant in time and are nearly the same for the three foam samples. For each foam sample, we calculate $K_o$ from Eq.~\ref{eq_avgvonNeumann}, or Eq.~\ref{eq_dAdt} for $p=1$ (since they are the same equation) using the slopes from part (a) as well as the individual values of $\langle A^2 \rangle / \langle A \rangle^2$ and $\langle \langle n \rangle \rangle$ for each foam; we find that $K_o=\left[0.022 \pm 0.001, 0.025 \pm 0.001  \right]$~mm$^2$/min where the first value is for Foam A and B, and the second value is for Foam C. These experimental values depend on the choice of gas and solution, which are the same for all three foams; however the larger value for $K_o$ for Foam C is likely due to the chemical aging of the foaming solution. But the data for Foam A and B, which were acquired one day apart and collapse, demonstrate repeatability. Taking all metrics together we establish the three foam samples are in the self-similar regime.

\begin{figure}[h]
\includegraphics[width=3.5in]{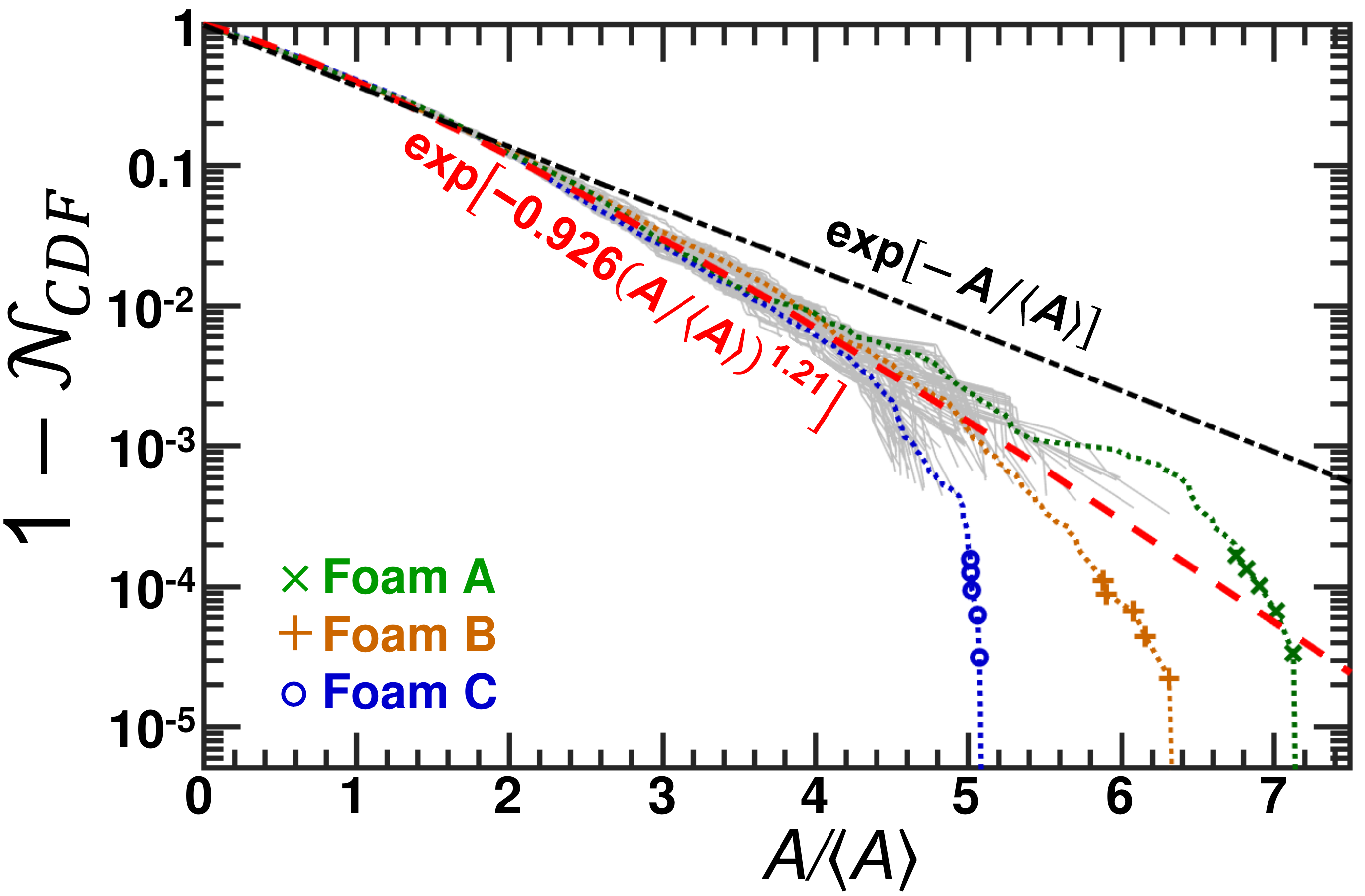}
\caption{Cumulative distribution function for bubble areas for the three coarsening samples, separately averaged over time and shown as dotted curves; the largest five observed bubble areas for each sample are plotted as symbols. Solid gray curves are distributions at different times, for three different foam samples, which contribute to the averages. The black dot-dashed curve is an exponential and the dashed curve is a compressed exponential.}
\label{CDF_a_norm}
\end{figure}

Finally, as an additional check on the data, Fig.~\ref{scaling_state_data}(c) shows average side number $\langle n \rangle \approx 6$ for each of the three samples.  By Euler's rule this should equal six for large enough samples, whether or not they are in the scaling state. We note that for each quantity in Fig.~\ref{scaling_state_data}(b) and (c) the values are nearly the same between foam samples. This along with the collapse of all of the area distributions between the three samples in Fig.~\ref{CDF_a_norm}, are further proof of good repeatability regardless of the physical chemistry of the foaming solution.

\begin{figure}[h]
\includegraphics[width=3.5in]{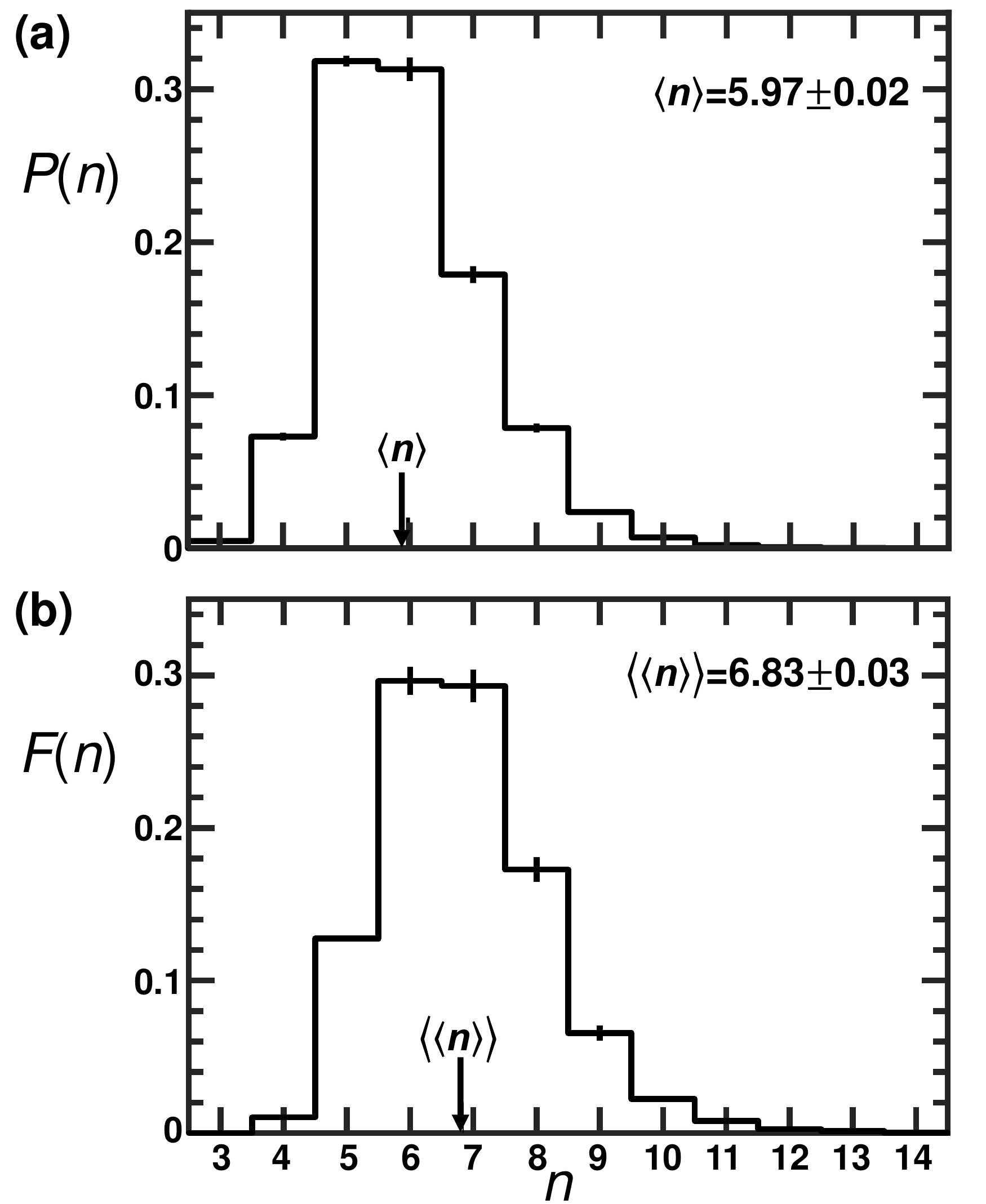}
\caption{(a) Side-number distribution and (b) area-weighted side-number distribution for the three coarsening foam samples, averaged together over time. The error bars span the values coming from the three different foam samples.  The average number side number $\langle n\rangle$ and the area-weighted average side number $\langle\langle n\rangle\rangle$ are indicated by arrows as labeled. } 
\label{p_of_n}
\end{figure}

Having established the foam is self-similar it follows that the distributions for side-number $P(n)$ and area-weighted side number $F(n)$ must also collapse throughout time. Therefore all of the data are averaged together for all times and for all three foam samples, giving one final distribution for $P(n)$ and one for $F(n)$; these are plotted in Fig.~\ref{p_of_n} parts (a) and (b), respectively. For both, the distributions are almost the same when comparing between experiments and the error bar span the calculated values for each $n$. Fig.~\ref{p_of_n}(a) shows that 5 and 6-sided bubbles are most prevalent. From $P(n)$, the average number of sides is computed to be $\langle n \rangle =5.97\pm0.02$. This is slightly smaller than 6, as expected, because the samples are finite.

Fig.~\ref{p_of_n}(b) shows that the average $F(n)$ distribution for the three foams is skewed more towards cells with large $n$ in comparison to $P(n)$, particularly for bubbles with $n=\{7,8\}$ sides. This is understood because bubbles with a larger number of sides tend have larger areas; the area-weighted average side number, $\langle\langle n \rangle \rangle = 6.83\pm0.03$, is therefore larger than $\langle n\rangle$.

Since the samples are self-similar, the cumulative area distributions (CDFs) must be independent of time when plotted versus $A / \langle A\rangle$. This is exemplified in Fig.~\ref{CDF_a_norm} for each of the three coarsening foams, where the gray curves are data from different times and where $y = 1-\mathcal{N}_{CDF}$ is the fraction of bubbles whose normalized areas are larger than $x=A/\langle A\rangle$. The gray curves collapse to the CDF made from all the normalized bubble areas collected into one distribution for each foam. The three total distributions also collapse and the data are found to follow the same slightly-compressed exponential as was found in \cite{RothPRE2013}, and is consistent with other works \cite{GlazierWeaireRev, StavansRev}. This behavior, while expected, is important because it demonstrates that all the normalized moments of the area distribution are independent of time; this fact is a main ingredient in both the derivation and expectations of the size-topology relations to be tested next.  Note, however, that the data have a cutoff where the CDF plummets to zero above the largest observed bubbles, whereas the compressed exponential fit has no such cutoff.  Therefore large-$p$ moments, which emphasize the largest bubbles, will have significant systematic differences that depend on sample size.

\subsection{Testing the Size-Topology Relations}

\begin{figure}[h]
\includegraphics[width=3.5in]{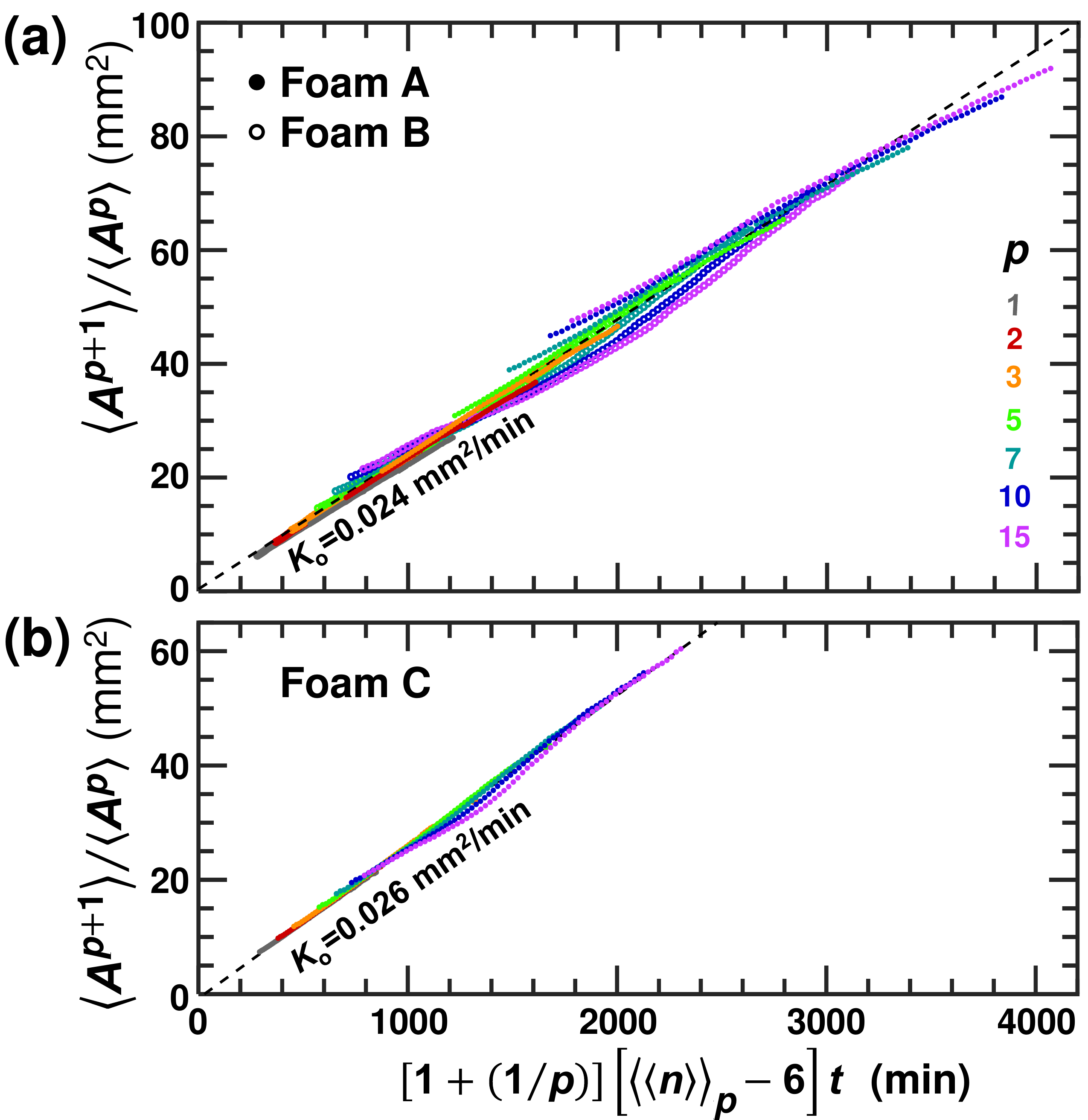}
\caption{Area moments normalized by preceding moment versus weighted time for various powers of $p$ as labeled.  Part (a) show Foam A and B because they have the same coarsening rate as determined by Fig.~\ref{scaling_state_data} and part (b) shows Foam C. Plotted as such, the data collapse and have proportionality constant $K_o$ taken from the Fig.~\ref{scaling_state_data}a analysis, in accord with the prediction of Eq.~\ref{eq_DAppOverAp}.}
\label{dynamic_test}
\end{figure}

We are interested in testing two of our generalized forms of Eq.~\ref{eq_avgvonNeumann}. The first is Eq.~\ref{eq_DAppOverAp}, relating the rate of change of two of the area moments to the area-weighted side number. Since the foam is self-similar the equation can be solved to show that $\langle A^{p+1}\rangle / \langle A^p \rangle$ grows linearly in time. By choosing any value of $p$ and evaluating the equation the data should collapse to a line with slope $K_o$.  This is done for several values of $p$ and Fig.~\ref{dynamic_test}(a) shows data for Foam~A and B, while Fig.~\ref{dynamic_test}(b) shows data for Foam~C since it has a different $K_o$ value per Fig.~\ref{scaling_state_data}a. For Foams A and B the data collapse to the same line with slope $K_o$ that is consistent with the values calculated from Fig.~\ref{scaling_state_data} data. Foam~C shows the best collapse of the three data sets and also has $K_o$ within error of the prior analysis.

\begin{figure}[h]
\includegraphics[width=3.5in]{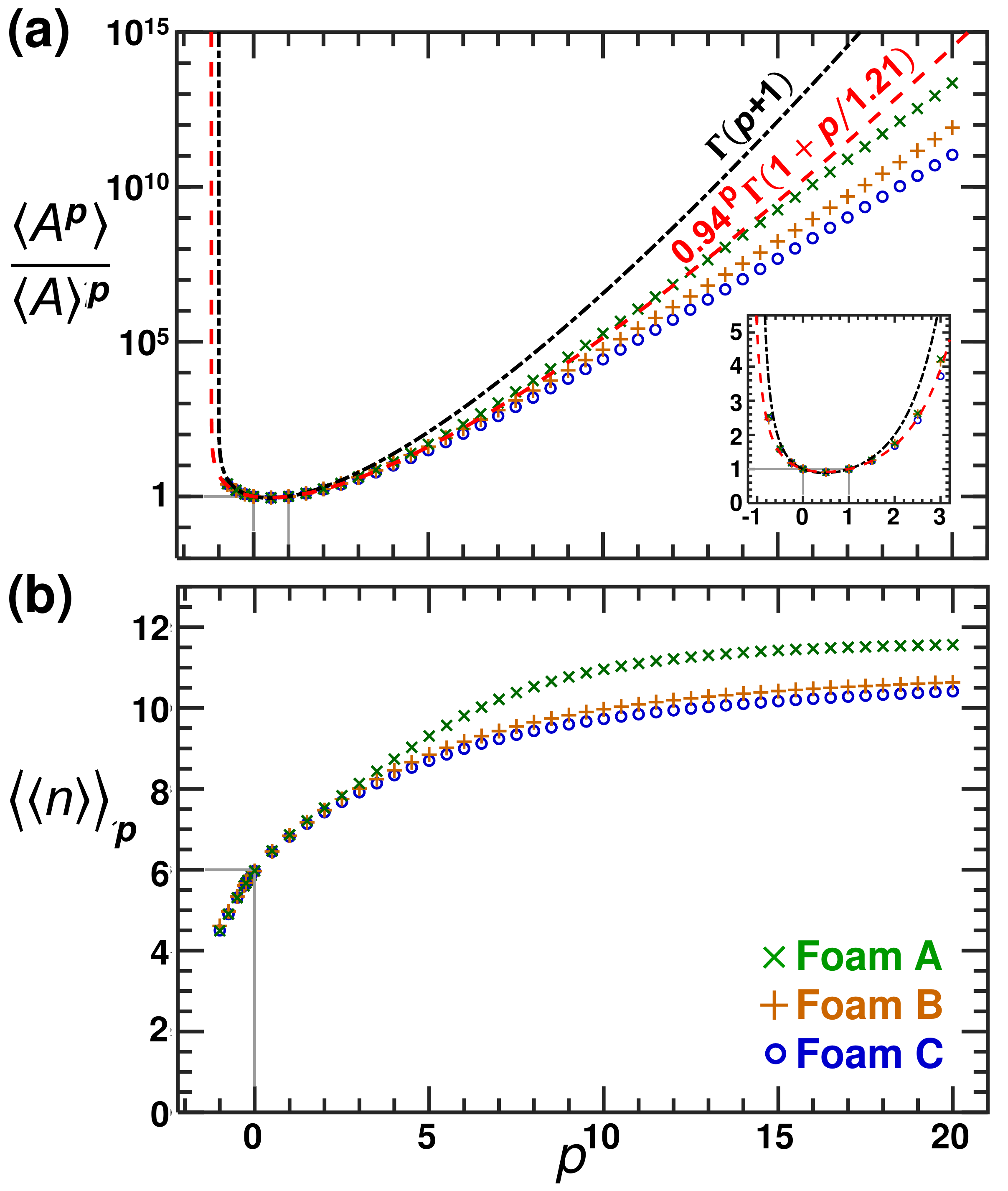}
\caption{Dimensionless moments of the area distribution (a) and weighted side number for areas raised to different powers (b) versus the power use in calculation. These quantities are calculated using data from all times for each of the three samples as labeled. Part (a) shows gamma functions as black dash-dotted and red dash curves that respectively correspond to the exponential and compressed exponential CDFs shown in Fig.\ref{CDF_a_norm}. The gray lines in both parts point to special cases for each quantity: part (a) they show where $\langle A^p \rangle / \langle A\rangle^p =1$ and part (b) shows where $\langle \langle n  \rangle \rangle_p=6$.   }
\label{raw_data}
\end{figure}

Eq.~\ref{eq_DAppOverAp} may be the cleanest looking of our generalizations but it is similar to Eq.~\ref{eq_avgvonNeumann} in that it is a rate of change. However, Eq.~\ref{eq_sizetopo} relates the sizes of the bubbles directly to their topology. There are two main ingredients in the equation and those are the dimensionless moments of the area distribution and the $A^p$-weighted side number. We evaluate these quantities using data from all times and they are individually plotted in Fig.~\ref{raw_data}(a) and (b), respectively. We note that while Eq.~\ref{eq_sizetopo} calls for values of $p \geq 2$ this is not the case of its components which can be calculated for any power. Therefore the $x$-axis in Fig.~\ref{raw_data} extends to negative numbers and we also evaluate the quantities for non integer values of $p$. 

One striking feature of the figure is that the data for Foam A appears separate from the data from Foam B and C; this is different than the coarsening behavior where Foam A and B are similar and Foam C is different. This separation does not happen until values of $p \gtrsim 7$ as evidenced by the inset in Fig.~\ref{raw_data}(a). The inset demonstrates how the data are well described by the moment generating function for the compressed exponential we fit to the area distribution. However the expectation deviates from the data for large $p$, likely due to finite size effects and/or because there is no cutoff for the assumed size distribution. It matches the data for Foam A best because it has the largest bubble(s) of all three data sets.  This is similar to what we see from Fig.~\ref{raw_data}(b) where the large-$p$ data for Foam A separates from the other two data sets. This too is explained by the largest bubble being in Foam A because large bubbles also have a large number of sides. Therefore for large-$p$ the calculation of $\langle \langle n \rangle \rangle_p$ is dominated by the bubbles with the most number of sides which also have very large areas. What is important for the size-topology equation though is how these quantities relate to the ones evaluated at a $p$ integer step above them and that they are time independent. We have already shown the latter and now want to evaluate the entirety of Eq.~\ref{eq_sizetopo} to see if the data follow the expectation.

\begin{figure}[h]
\includegraphics[width=3.5in]{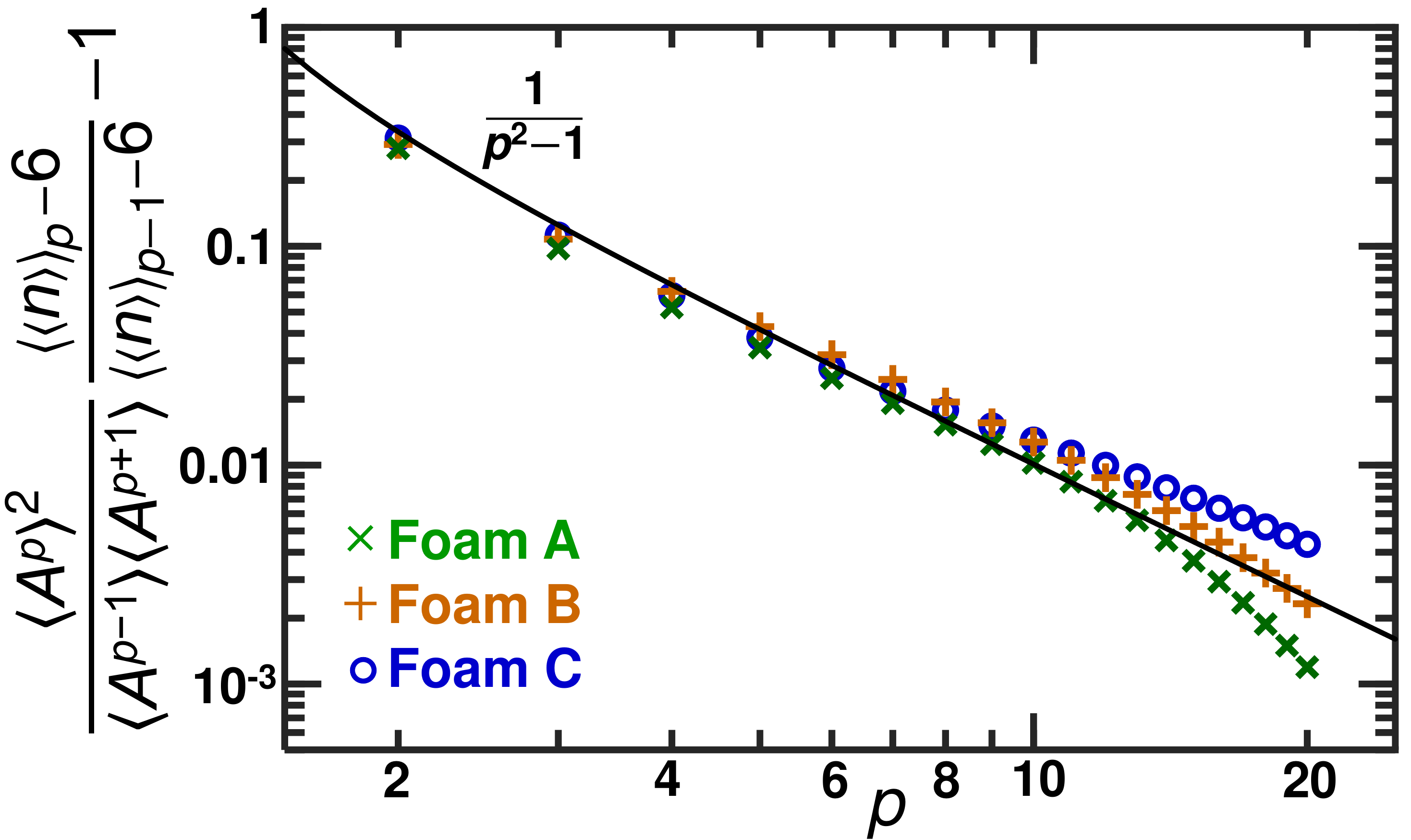}
\caption{The generalized size-topology identity versus power. The different symbols represent three different coarsening foams. The data collapse to the expectation which is plotted as a black curve.}
\label{final_identity}
\end{figure}

Using the various dimensionless moments of the area distribution and the $A^p$-weighted average side number we now compute the size-topology relationship of Eq.~\ref{eq_sizetopo}. It is evaluated for $2 \leq p \leq 20$; the expectation asymptotes to 1 so the resulting values have $1$ subtracted from them and are plotted in Fig.~\ref{final_identity} for each of our three foams. Performing this subtraction along with plotting the $y$-axis on a log-scale is done to get a closer look at the data.  This reveals good agreement with the expectation. However, there are deviations at large $p$ but they are explained by finite size effects and the sensitivity of large-$p$ moments to the largest bubbles in the sample.

\section{Conclusion}

We have derived several identities for 2d foams in a self-similar scaling state. The ones we tested experimentally are: Eq.~\ref{eq_DAppOverAp}, which compares the rate of change of successive moments of the area distribution to the $A^p$-weighted average side number; Eq.~\ref{eq_sizetopo}, which is a generalized size-topology relation that relates moments in the area distribution to the $A^p$-weighted side number. Both equations are derived using only von~Neumann's law and the fact that the foams are in a self-similar scaling state. We tested these relationships for three different foam samples. After showing the foam is self-similar we found the data agree well with both predictions. A natural extension would be to study 3d foams; however, since the von~Neumann like expression for domain growth in 3d is not purely topological \cite{MacPhersonSrolovitz2007}, any size-topology relationships would be approximate. Instead, future research might focus on the dynamic size-topology equation that was derived here in the form of Eq.~\ref{eq_dAp} but was not tested. Another avenue for future work would be to consider how the size-topology relations constrain the form of the bubble size distribution and if they could permit it to be derived. It would be particularly interesting to investigate if there is a cutoff in the distribution, \emph{i.e.}~if there is a maximum possible bubble size in comparison with the average. Further work could also explore how $d\langle A^p\rangle/dt$ is affected by nonzero wetness, where von~Neumann's law must be modified to account for transport across the Plateau borders as predicted in \cite{SchimmingPRE2017}, and tested in \cite{ATCcoarsen}; this is challenging because the corrections to von~Neumann depend not only on the size of the Plateau borders but also on the shapes of the bubbles.


\section*{Author Contributions}

All authors conceived and designed the study.  DJD performed the calculations, while ATC performed the experiments and analyses with guidance by DJD and JPS. DJD and ATC wrote the manuscript with input from JPS.

\section*{Funding}
This work was supported by NASA grant 80NSSC19K0599.



\bibliography{CombineRefs}

\end{document}